# Fluctuating Fractionalized Spins in Quasi Two-dimensional Magnetic $V_{0.85}PS_3$

Vivek Kumar[1,*], Deepu Kumar[1], Birender Singh[1], Yuliia Shemerliuk[2,] Mahdi Behnami[2], Bernd Büchner[2,3], Saicharan Aswartham[2], and Pradeep Kumar[1,†]

[1] *School of Basic Sciences, Indian Institute of Technology Mandi, Mandi-175005, India*
[2] *Leibniz-Institute for Solid-state and Materials Research, IFW-Dresden, 01069 Dresden, Germany*
[3] *Institute of Solid-State Physics, TU Dresden, 01069 Dresden, Germany*

**Abstract**

Quantum spin liquid (QSL), a state characterized by exotic low energy fractionalized excitations and statistics is still elusive experimentally and may be gauged via indirect experimental signatures. Remnant of QSL phase may reflect in the spin dynamics as well as quanta of lattice vibrations, i.e., phonons, via the strong coupling of phonons with the underlying fractionalized excitations i.e., Majorana fermions. Inelastic light scattering (Raman) studies on $V_{1-x}PS_3$ single crystals evidences the spin fractionalization into Majorana fermions deep into the paramagnetic phase reflected in the emergence of a low frequency quasielastic response along with a broad magnetic continuum marked by a crossover temperature $T^*$ ~ 200 K from a pure paramagnetic state to fractionalized spins regime qualitatively gauged via dynamic Raman susceptibility. We further evidenced anomalies in the phonons self-energy parameters in particular phonon line broadening and line asymmetry evolution at this crossover temperature, attributed to the decaying of phonons into itinerant Majorana fermions. This anomalous scattering response is thus indicative of fluctuating fractionalized spins suggesting a phase proximate to the quantum spin liquid state in this quasi two-dimensional (2D) magnetic system.

[*]vivekvke@gmail.com
[†]pkumar@iitmandi.ac.in

**INTRODUCTION**

The family of transition metal phosphorous tri-chalcogenides (*TM*PX$_3$, *TM* = V, Mn, Fe, Co, Ni or Zn and X = S, Se) with a strong in-plane covalent bonds and weak van der Waals gap between the layers of magnetic atoms have appeared as an intriguing candidate for exploring the quasi-2D magnetism, where the inter-planar direct- and super-exchange magnetic interactions are substantially quenched [1-3]. The underlying magnetic ground state ($|GS\rangle$) in these materials is affected by the trigonal distortion (*TM*X$_6$) due to change in the local symmetry from $O_h$ to $D_{3d}$, which consequently lifts the degeneracy of the *d*-orbitals. Magnetic studies on these materials reveals the emergence of quite different magnetic $|GS\rangle$ on varying *TM* atoms; such as, FePS$_3$ shows an Ising-type transition at $T_N$ ~ 123 K, NiPS$_3$ /MnPS$_3$ undergoes a *XY*/ Heisenberg-type transition at $T_N$ = 155 K / 78 K, respectively [4]. We note that in this *TM*PX$_3$ family, vanadium-based system remains largely unexplored beyond their basic properties. The recent reports on V$_{0.9}$PS$_3$ revealed an insulator-to-metal phase transition at very nominal pressure (~ 26 Kbar) without any structural transition and realized the Kondo effect within the metallic phase. It was also advocated that this system lies in the close proximity of the quantum spin liquid (QSL) state [1]. As the insulator-to-metal transition is also associated with antiferromagnetic (AFM) to paramagnetic transition, and opens the possibility of a highly entangled spin liquid phase due to honeycomb lattice via potential Kitaev interactions. Interestingly, recently a new kind of Kondo behavior has been proposed and is attributed to gauge fluctuations from bond defects in spin-liquids [5]. These experimental observations and theoretical predictions suggest a key route to the observation of fascinating QSL state in these quasi 2D quantum magnetic materials. An important characteristic of these systems, is that despite all being isostructural, magnetic lattice is 2D honeycomb structure formed by TM ions, have different spin dimensionality. For example, $|GS\rangle$ dynamics of MnPS$_3$ member of this family is described by isotropic ($J_\perp = J_\parallel$) Heisenberg Hamiltonian [$H = -\sum_{ij} J_\perp (S_{ix}.S_{ix} + S_{iy}.S_{iy}) + J_\parallel S_{iz}.S_{iz}$]; FePS$_3$ using Ising ($J_\perp = 0$) model, whereas NiPS$_3$ is understood using anisotropic ($J_\perp > J_\parallel$) model. This complexity is further enhanced in the $|GS\rangle$ of V-based system, V$_{0.85}$PS$_3$, also reflected in the magnetic measurements [6], where the magnetic susceptibility exhibits intriguingly different behavior compared with other members of this family, suggesting the presence of an additional exotic competing interaction in the Hamiltonian, such as Kitaev, to completely understand the underlying magnetic $|GS\rangle$.

Patterns of the exotic $|GS\rangle$ magnetic properties in this system may emerge from the presence of strong quantum fluctuations due to entanglement of the underlying spins within the honeycomb lattice i.e., a QSL state. Interest in the field of QSL was renewed with the seminal work of Kitaev in 2006 [7] and subsequently certain conditions were laid down for the realization of a QSL state such as [8] presence of hexagonal honeycomb lattice, mott-insulator, edge sharing octahedra in the structure and spin-orbit coupling. A large number of systems have been proposed, however so far there is not a single system which perfectly display QSL state as a true $|GS\rangle$. Although numerous proposed systems do show strong signature of a QSL state or a proximate QSL state [9-11]. However, in all those cases an ideal QSL state is preempted by the long-range magnetic ordering at low temperature. Despite this, the signature of a QSL state may be captured as fluctuation in the short-range ordering regime much before setting up of the long-range ordering. Recently, it was shown that in a magnetic system, CrSi/GeTe$_3$, with $S = 3/2$ and very weak spin-orbit coupling does reflect the signature of a QSL state understood using $S = 3/2$ XXZ-Kitaev model given as

$$H_{XXZ-K} = \sum_{\langle i,j \rangle} \frac{J_\perp}{2} (S_i^+ S_j^- + S_i^- S_j^+) + J_\parallel S_i^z S_j^z) + K \sum_{\langle i,j \rangle \alpha} S_i^\alpha S_j^\alpha$$ [12,13]. Recently, signature of

fractionalized excitations was also reported in a magnetic $S=1/2$ Kagome AFM systems attributed to a remnant QSL state [14]. Also, in case of a putative QSL candidate α-RuCl$_3$, it was shown that the low temperature zigzag AFM state is stabilized by quantum fluctuations with spin liquid state as a proximate metastable state [15]. Generally, for a quantum spin system frustration, dimerization, interchain/interlayer coupling, vacancy/defects leading to bond disordered, and spin-phonon coupling have an impact on the dynamics of the low energy excitations [16-18]. In 2D quantum systems, spin liquid $|GS\rangle$ is naturally a consequence of exotic topologies such as hexagonal honeycomb structure as both the triangular and square lattice have AFM like $|GS\rangle$ [16]. For V$_{1-x}$PS$_3$, we do have *p-p* dimer formation in the *b-c* plane (see Fig. 1), vacancy, hexagonal honeycomb lattice, hinting that it may have remnant of QSL state as strong quantum fluctuations. Inherent coordination flexibility and electronic configuration of V ion makes it possible to realize exotic exchange with topologies.

Motivated by these diverse suggestions and possibilities of a QSL state in these quasi 2D magnetic systems, we carried out an in-depth Raman scattering studies on single crystal of V$_{0.85}$PS$_3$ to understand the underlying exotic properties. Evidences of a QSL state or its remnant may be uncovered via observation of the quantum fluctuations of the associated spin degrees of freedom and their coupling with the lattice degrees of freedom through spin-phonon coupling. Inelastic light (Raman) scattering is an excellent technique to probe such dynamic quantum fluctuations reflected via the emergence of the quasielastic response at low energy and a broad continuum in the Raman response $\chi''(\omega,T)$ [19-23], smoking gun evidence of a QSL state. Here, we observed a strong low energy quasielastic response with lowering temperature and a broad continuum; quite startling, it starts emerging much above the long-range magnetic ordering temperature. These characteristic features clearly suggest the presence of strong underlying quantum fluctuations.

Surprisingly, the corresponding estimated dynamic Raman susceptibility, $\chi^{dyn}$, amplitude is not quenched below $T_N$; as expected for a conventional magnetically ordered system, signalling that it emerges from a proximate QSL state or its remnant. Our observations evince the signature of a remnant QSL state as fluctuating part suggests that this system lies in proximity of a QSL ground state. This also suggest that the low temperature ordered phase may be proximate to the quantum phase transition into a spin liquid $|GS\rangle$. The anomalies observed in the $\chi^{dyn}$ maps parallel with the anomalies seen in the self-energy parameters of the optical phonons i.e., peak frequencies and line-widths. Here, we report the experimental evidence supporting the existence of a remnant QSL phase in $V_{0.85}PS_3$ using Raman spectroscopy.

**RESULTS AND DISCUSSION**

**Temperature evolution of the broad magnetic continuum - Fractionalized fermionic excitations**

Magnetic Raman scattering give rise to a broad continuum originates from underlying dynamical spin fluctuations and may be used to gauge the fractionalization of quantum spins expected for proximate spin liquid candidates [24-31]. To investigate possible emergence of the fractionalized fermionic excitations, we carried out a detailed analysis of the temperature evolution of the observed broad magnetic continuum in the Raman spectra. The fractional spin excitations play a key role in dictating the temperature evolution of the background continuum of the Raman spectra because their occupation is determined by the Fermi distribution function. First, we focus on the temperature evolution of the integrated intensity of the background continuum to determine the fractionalized excitations energy scale [29], where the raw Raman intensity $I(\omega)$ is integrated over a range of 1.0 meV - 95 meV as: $I = \int_{\omega_{min}}^{\omega_{max}} I(\omega,T)\,d\omega$. Figure 2(d) shows the temperature evolution of the integrated intensity of the background continuum. As can be seen from the Fig. 2(d), the integrated intensity of the background continuum shows a non-monotonic temperature dependence, at high temperature the intensity variation is mainly dominated by a conventional one particle scattering corresponding to thermal Bose factor given as: $I(\omega,T) \propto [1+n(\omega,T)]$; where $[1+n(\omega,T)] = 1/[1-e^{-\hbar\omega_b/k_BT}]$. However, at low temperature a significant deviation from the conventional bosonic excitations is observed below ~150 K, the intensity shows monotonic increase with decreasing temperature down to lowest recorded temperature (4K). To understand the temperature dependence of the integrated intensity of the background continuum, we fitted with a function having contribution from both Bosonic and two-fermion (related to the creation and annihilation of the pair of fermions, its functional form is given as: $a+b[1-f(\omega,T)]^2$, where $f(\omega,T) = 1/[1+e^{\hbar\omega_f/k_BT}]$ is Fermi distribution function with zero chemical potential [29]), excitations. The fitting outcome reveals that the temperature dependence of the integrated intensity below ~ 150 K is mainly dominated by the fermionic excitations and the corresponding fractionalized energy scale for the fermions is $\omega_f$ = 10.3 meV (~ 85 cm$^{-1}$). The temperature evolution of the integrated intensity of Bose-subtracted spectra is shown in Fig. 2(d), clearly indicating the significant enhancement of magnetic contribution below ~ 150-200 K, as fitting well with the two-fermion function, $a+b[1-f(\omega,T)]^2$. Our analysis clearly evidences the signature of fractionalized fermionic excitations in the V$_{0.85}$PS$_3$, a key signature of proximate QSL phase, inline with the other 2D honeycomb putative QSL candidates, $\alpha$-RuCl$_3$ as well as Li$_2$IrO$_3$ [28,30].

Now, we will discuss the Raman response, $\chi''(\omega,T)$, which shows the underlying dynamic collective excitations at a given temperature, where $\chi''(\omega,T)$ is calculated by dividing raw Raman intensity with Bose factor, $I(\omega,T) \propto [1+n(\omega,T)] \chi''(\omega,T)$. The Raman response, $\chi''(\omega,T)$, is proportional to stokes Raman intensity given as: $I(\omega,T) = \int_0^\infty dt\, e^{i\omega t} \langle R(t)\, R(0) \rangle \propto [1+n(\omega,T)]\chi''(\omega,T)$; where $R(t)$ is the Raman operator and $[1+n(\omega,T)]$ is the Bose thermal factor. Figure 2(a) shows the temperature evolution of the $\chi''(\omega,T)$. We note that $\chi''(\omega,T)$ is composed of the phononic excitations superimposed on a broad continuum extending up to ~ 95 meV. The detailed analysis of the self-energy parameters of all the observed phonon modes i.e. peak frequency and linewidth is given in the supplementary information [6]. An in-depth analysis of this broad continuum may provide the further information about the underlying nature of the dynamical spin fluctuations via the dynamic Raman susceptibility ($\chi^{dyn}$). Interestingly the Raman response shows a significant increase in the spectral weight on lowering the temperature [see Fig. 2(a) and its inset], and quite surprisingly it continues to increase upon entering into the spin solid phase unlike the conventional systems where it is expected to quench below $T_N$. This characteristic scattering feature is typical of the scattering from underlying quantum spin fluctuations. For further probing the evolution of this broad magnetic continuum and underlying quantum spin fluctuations we quantitatively evaluated $\chi^{dyn}$, shown in Fig. 2 (c). $\chi^{dyn}$ at a given temperature is evaluated by integrating phonon subtracted Raman conductivity, $\chi''(\omega)/\omega$, shown in figure 2 (b), and using Kramers - Kronig relation as:

$$\chi^{dyn} = \lim_{\omega \to 0} \chi(k=0,\omega) \equiv \frac{2}{\pi}\int_0^\Omega \frac{\chi''(\omega)}{\omega}d\omega \qquad (1)$$

where $\Omega$ is the upper cutoff value of integrated frequency chosen as ~ 95 meV, where Raman conductivity shows no change with further increase in the frequency. With lowering temperature $\chi^{dyn}$ shows nearly temperature independent behavior down to ~200 K as expected in a pure paramagnetic phase, on further lowering the temperature it increases continuously till 4K. In the quantum spin liquid phase, Raman operator couples to the dispersing fractionalized quasi particle excitation and reflects the two-Majorana fermion density of states [26]. Therefore, increase in the $\chi^{dyn}$ below ~ 200 K reflects the enhancement of Majorana fermion density of states and marks the cross over from a paramagnetic to the proximate spin liquid state where fractionalized excitations start building up. Remarkably, the temperature dependence of the phonon modes also showed the anomalies around ~ 200 K, reflecting the strong coupling of fractionalized excitations with the lattice degrees of freedom (discussed in later section). For conventional antiferromagnets as system attains ordered phase dynamical fluctuations should be quenched to zero, contrary here we observed a significant increase in dynamic Raman susceptibility hinting strong enhancement of dynamic quantum fluctuations [16,31-33]. The diverging nature of $\chi^{dyn}(T)$ as T → 0K, clearly suggests the dominating nature of quantum fluctuations associated with the underlying collective excitations down to the lowest temperature. This is also consistent with recent theoretical understanding, where it was advocated, that dynamic correlations may have unique temperature dependence in system with quantum spin liquid signatures and the fractionalization of the quantum spins contribute to dynamic spin fluctuations even in the high temperature paramagnetic phase [34]. Therefore, naturally the signature of spin fractionalization is expected to be visible in the dynamical measurable properties such as dynamic Raman susceptibility as observed here. It was also shown that in the low temperature regime dynamical structure factor, related with spin-spin correlation function, shows quasielastic response with lowering temperature and was suggested as evidence for fractionalization of spins.

Next, we focus on the very low frequency region (LFR) i.e., 1-9 meV, where we observed the emergence of a strong quasielastic response at low temperature, see Fig. 3 (a and b). We evaluated the dynamic Raman susceptibility, $\chi^{dyn}_{LFR}$ for this low energy range (see Fig. 3 (c)). The observed $\chi_{LFR}^{dyn}$ remains nearly constant till ~ 200 K, and shows monotonic increase with further decreasing the temperature down to 4K. We fitted both $\chi^{dyn}$ and $\chi_{LFR}^{dyn}$ using a power law as $\chi^{dyn} \propto T^\alpha$ ($\alpha =$ -0.34 for $\chi^{dyn}$, see solid blue line in Fig. 2 (c); and $\alpha =$ -0.67 for $\chi_{LFR}^{dyn}$, see solid line in Fig. 3 (c)). Here we observed power law behavior of $\chi^{dyn}$ and $\chi_{LFR}^{dyn}$ much above $T_N$ unlike the conventional pure paramagnetic phase where it is expected to show saturation. The power law dependence of $\chi^{dyn}$ and $\chi_{LFR}^{dyn}$ even well above the long-range magnetic ordering temperature reflects the slowly decaying correlation inherent to the quantum spin liquid phase and triggers fractionalization of spins into itinerant fermions at T* ~ 200 K [35,36]. We note that this anomalous temperature evolution of the background continuum along with phonon anomalies, discussed later, cannot be captured by the conventional long-range ordered magnetic scattering, rather, it reflects the presence of fractionalized excitations which are intimately linked with the quantum spin liquid phase, in line with the theoretical suggestions for a QSL state.

We wish to note that in a recent report for the case of a putative QSL candidate RuCl$_3$ [15], QSL phase is predicted with long range ordering at T$_N$ ~ 7K (Zigzag AFM state). It is shown that in the high temperature paramagnetic phase quasielastic intensity of magnetic excitation have broad continuum and the low temperature AFM state is quite fragile with competition from FM correlation and QSL phase; in fact, AFM state is advocated to be stabilized by Quantum fluctuations leaving QSL and FM states as proximate to the $|GS\rangle$ and at slightly higher temperature Kitaev QSL state becomes prominent. Furthermore, it was advocated that FM and QSL states proximate to AFM $|GS\rangle$ are essential to understand the anomalous scattering continuum. Based on our observations and phonon anomalies (discussed in next section) this broad magnetic continuum is attributed to the fractionalized fermionic excitations in this material with quasi 2D magnetic honeycomb lattice.

**Mode's asymmetry and anomalous phonons**

Interaction of the underlying magnetic continuum with the lattice degrees of freedom may reflect via asymmetric nature of the phonon line shape, known as Fano-asymmetry, and may provide crucial information about the nature of underlying magnetic excitations responsible for the magnetic continuum. This asymmetry basically describes the interaction of a continuum with a discrete state (optical-phonons here) and this effect have its origin in the spin-dependent electron polarizability involving both spin-photon/phonon coupling [37-39]. For the spin liquid candidates, recently it was advocated that spin-phonon coupling renormalizes phonon propagators and generates Fano line shape resulting into observable effect of the Majorana fermions and the $Z_2$ gauge fluxes, a common denominator for a quantum spin liquid $|GS\rangle$ [40,41]. The evolution of phonon modes asymmetry in putative QSL candidate materials seems ubiquitous [30,42-44] suggesting the intimate link between QSL phase and phonons asymmetric line shape. Additionally, it was also shown that the lifetime of the phonons decreases with decreasing temperature i.e., increase in linewidth with decreasing temperature, attributed to the decay of phonon into itinerant Majorana fermions [45]. This is opposite to the conventional behavior where phonon linewidth decreases with decreasing temperature owing to reduced phonon-phonon interactions.

Figure 4 (a) shows the temperature dependent Raman spectra in a frequency range where couple of modes (P2, P6 and P7) show strong asymmetric line evolution. It's very clear that these phonon modes gain asymmetric line shape with decreasing temperature. Interestingly these modes are also superposed on the broad underlying magnetic continuum in the region where spectral weight of the continuum is dominating. Asymmetric nature of these three phonons mode is gauged via slope method, see supplementary information (section S7) for details and equivalence between slope method and Fano function ( $F(\omega) = I_0(q+\varepsilon)^2/(1+\varepsilon^2)$; where $\varepsilon = (\omega-\omega_0)/\Gamma$ and $1/q$ defines as the asymmetry); here we have adopted a slope method because Fano function fitting resulted into large error. Figure 4 (b) shows the normalized slope for these three modes. Interestingly, the asymmetry gauged via slopes shows strong temperature dependence [see Fig. 4(b)]; it has a high value in the long-range-ordered phase at low temperature, above $T_N$ (∼60 K) it continuously decreases till ∼200 K, and thereafter it remains nearly constant up to 330 K, clearly suggesting the presence of active magnetic degrees of freedom far above $T_N$. A pronounced feature of this mode asymmetry is that it conjointly varies with dynamic Raman susceptibility on varying temperature [discussed above; see Fig. 2 (c) and 3 (c)], implying that its asymmetric line shape is also an indicator of spin fractionalization or emergence of spin liquid phase, and the increased value below 200 K ($T^*$- defined as a crossover temperature) may be translated to a growth of finite spin fractionalization. We also tried to fit the temperature evolution of the slopes with the two-fermion scattering form $a+b[1-f(\omega)]^2$, (see Fig. 4 (b)) and the extracted fermionic energy scale is also found to be similar to that estimated from intensity fitting of the continuum background (see Fig. 2 (d)). It is quite interesting that evolution of mode's asymmetry for these optical-phonons maps parallel to the thermal damping of the fermionic excitations. In materials with QSL phase as $|GS\rangle$, spins are fractionalized into the Majorana fermions, as a result of this the underlying continuum emerging from spin fractionalization couples strongly with lattice degrees of freedom as evidenced here. The temperature evolution of these modes, i.e., phonon mode's line asymmetry, in line with the theoretical prediction clearly evidenced the fractionalization of spins into Majorana fermions.

For a normal phonon mode behaviour, as the temperature is lowered then phonon peak energy is increased and linewidth decreases attributed to the anharmonic phonon-phonon interactions. Interestingly, large number of modes showed change in phonon frequency at the cross over temperature $T^*$ (~ 200K) signaling the effect of spin fractionalization on the optical phonons, see Fig. 3 (d, e and f). Startlingly, some of the modes showed anomalous evolution of the linewidth below $T^*$ i.e., linewidth increases with decreasing temperature. Fig. 4(c, d and e) shows the temperature dependence of the linewidth of the phonon modes which show anomalous behaviour. All these modes show clear divergence from the normal behaviour starting at the crossover temperature $T^*$, implying additional decay channel, similar to the temperature scale associated with the phonon mode's line asymmetry and magnetic continuum reflected via dynamic Raman susceptibility. This is also consistent with the theoretical predictions; hence our observation clearly evidenced the emergence of fractionalised excitations in this quasi 2D magnetic system starting from the crossover temperature reflected via phonon anomalies and broad underlying magnetic continuum.

**CONCLUSION**

In conclusion, we have performed in depth inelastic light scattering (Raman) studies on $V_{1-x}PS_3$ single crystals. Where, we focused on the background continuum showing distinct temperature dependence via dynamic Raman susceptibility and the phonons anomalies. Our results on background continuum and phonons self-energy parameters evince anomaly at the similar temperature range suggesting the cross over from a normal paramagnetic phase to a state where spin fractionalization begins and marked the onset of proximate quantum spin liquid phase. Our studies evinced the signature of spin fractionalization in this quasi 2D magnetic honeycomb lattice system. In addition to the observation of a broad magnetic continuum and its anomalous temperature evolution, our results on the evolution of the mode's line asymmetry and phonon anomalies, in particular for the phonon modes lying on the underlying magnetic continuum, opens the possibility to experimentally identifying the theoretically predicted effects of fractionalized excitations of QSL phase in putative spin liquid candidates.

**Acknowledgments**

P.K. thanks IIT Mandi and Department of Science and Technology - India for the financial support. S.A. acknowledges the support of Deutsche Forschungsgemeinschaft (DFG) through Grant No. AS 523/4–1. B.B. acknowledge financial support from the DFG through the Würzburg-Dresden Cluster of Excellence on Complexity and Topology in Quantum Matter – ct.qmat (EXC 2147, project-id 390858490).

**Figure Captions:**

**Figure 1:** Crystal structure of VPS$_3$ plotted using (VESTA) as viewed along **(a)** arbitrary direction. **(b)** *a*-axis, **(c)** *b*-axis and **(d)** *c*-axis, thick grey solid lines connecting vanadium atoms represent 2D honeycomb lattice formation. Red, blue and green color spheres represent Vanadium, Phosphorus and Sulphur, respectively.

**Figure 2: (a)** Temperature evolution of the Raman response $\chi''(\omega,T)$ [measured raw Raman intensity/$1+n(\omega,T)$]. Inset shows the phonon's subtracted Raman response. Lables P1-P15 represent phonon modes. **(b)** Temperature dependence of the phonon's subtracted Raman conductivity $\chi''(\omega,T)/\omega$. **(c)** Temperature dependence of $\chi^{dyn}(T)$ obtained from Kramer - Kronig relation. The solid blue line is power-law fit $\chi^{dyn} \propto T^\alpha$. Background colored shading reflects different magnetic phases. **(d)** Shows the magnetic contribution to the integrated intensity in the energy range 1 to ~ 95 meV, after subtracting the bosonic background (shown in the inset as green color line). Red solid line represents fitting by the two-fermion scattering function $a+b[1-f(\omega,T)]^2$; where $f(\omega,T)=1/[1+e^{\hbar\omega_f/k_BT}]$ is the Fermi distribution function. Inset shows bosonic and fermionic behavior of integrated intensity. $T^*$ (~ 200K) represents the temperature where spins fractionalization starts building up.

**Figure 3: (a)**, **(b)** Shows temperature evolution of the Bose-corrected spectra i.e. Raman response, $\chi''(\omega,T)$, and Raman conductivity, $\chi''(\omega,T)/\omega$, in the low frequency region LFR (1-9.0 meV), respectively. **(c)** Temperature dependence of $\chi^{dyn}_{LFR}(T)$ obtained from Kramer's Kronig relation. The solid blue line is power-law fit $\chi^{dyn}_{LFR} \propto T^\alpha$. Background colored shading reflects different magnetic phases. **(d, e and f)** Mode's frequency evolution as a function of temperature for the modes P2, P4, P6, P7, P9 and P10. Red solid lines are guide to the eye. $T^*$ (~ 200K) represents the temperature where spins fractionalization starts building up.

**Figure 4: (a)** Shows the raw Raman spectrum in the frequency range of ~ 45 – 360 cm$^{-1}$ showing the evolution of the asymmetry for the modes P2, P6 and P7. **(b)** Shows the evolution of the phonons modes P2, P6 and P7 (as inset) normalized asymmetry gauged via slope. Background colored shading reflects different magnetic phases. **(c, d and e)** Shows the linewidth for the modes P2, P4, P6, P7, P9 and P10. Red solid lines are guide to the eye. $T^*$ (~ 200K) represents the temperature where spins fractionalization starts building up.

**Figure 1:**

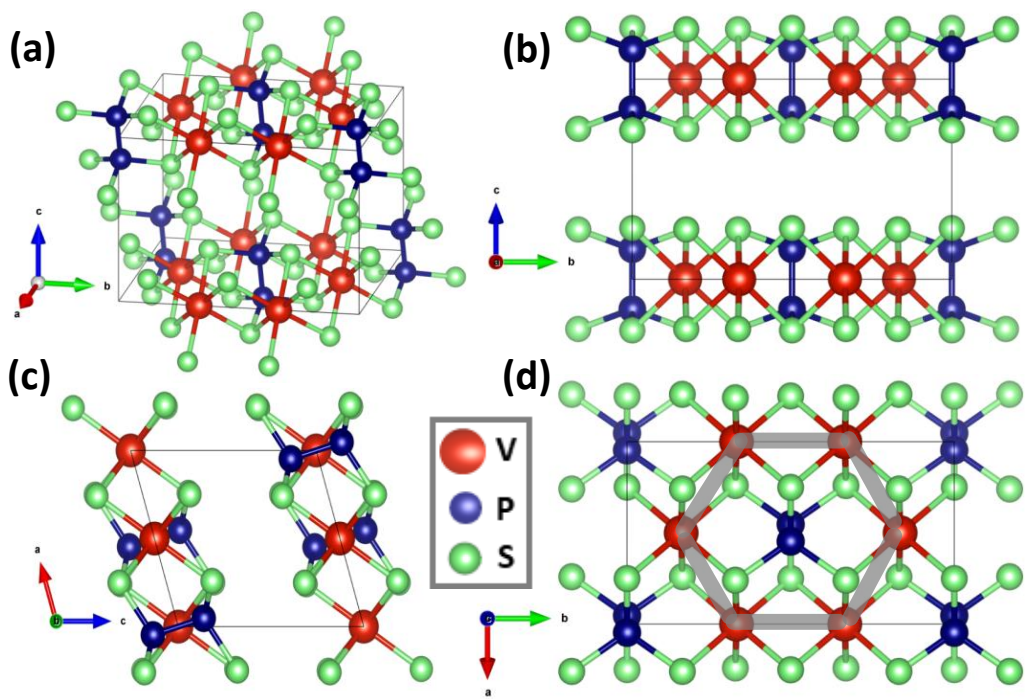

**Figure 2:**

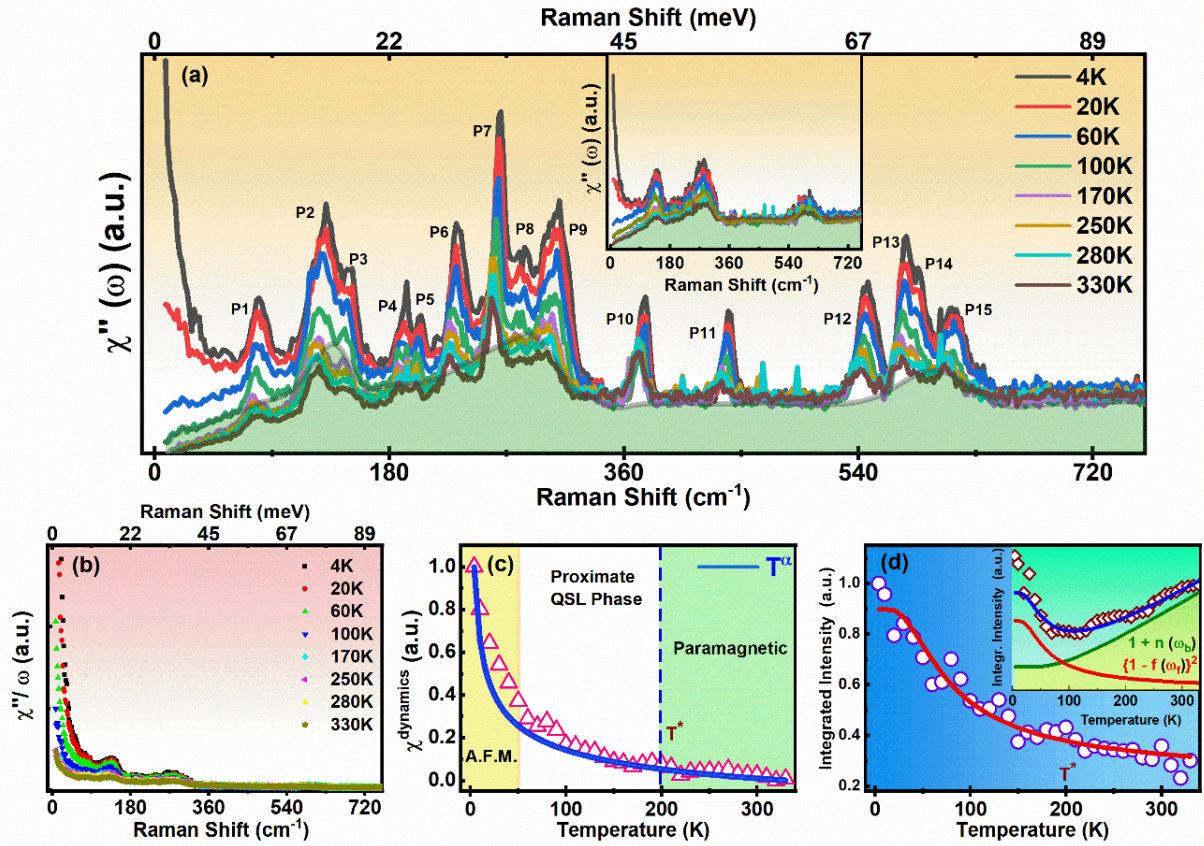

**Figure3:**

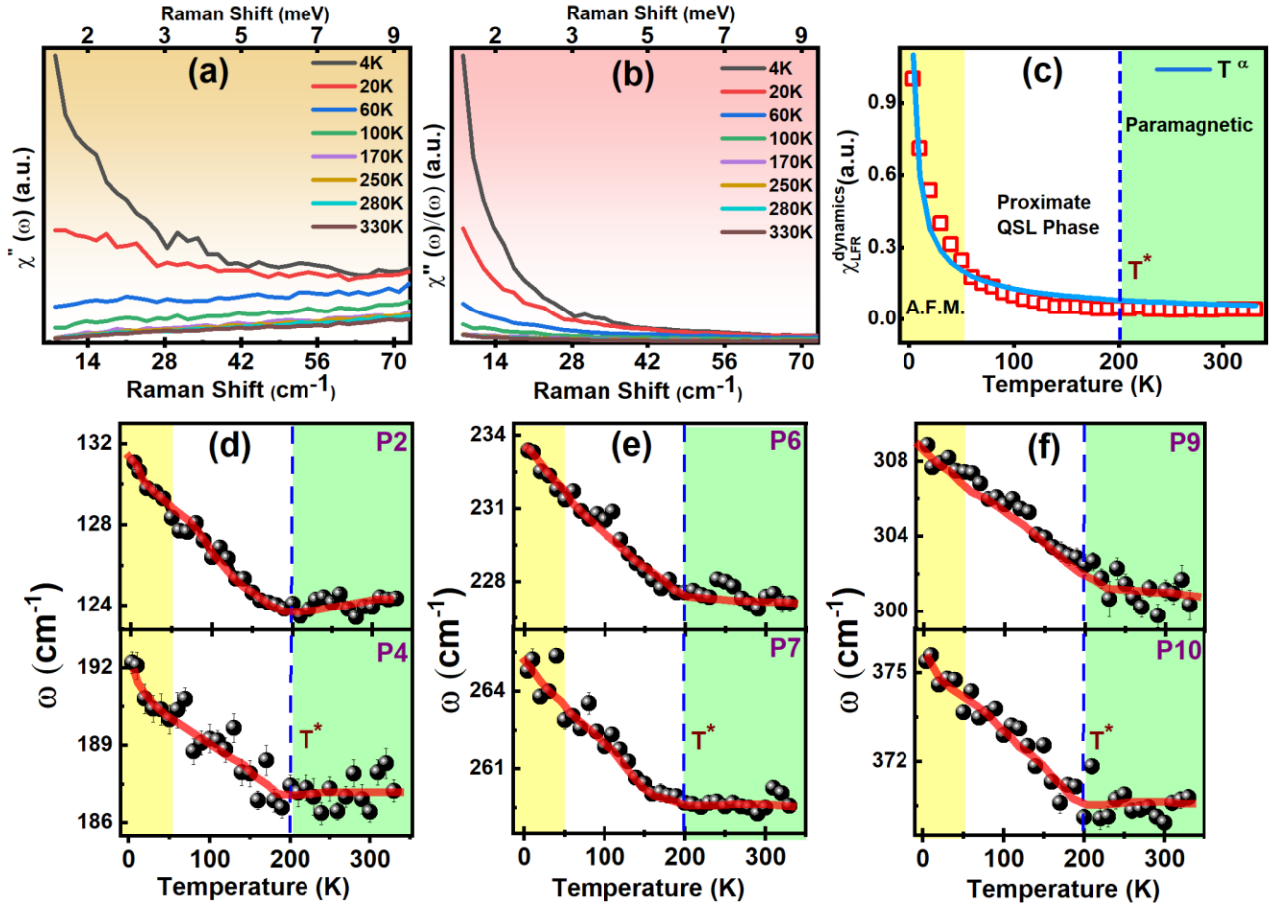

**Figure 4:**

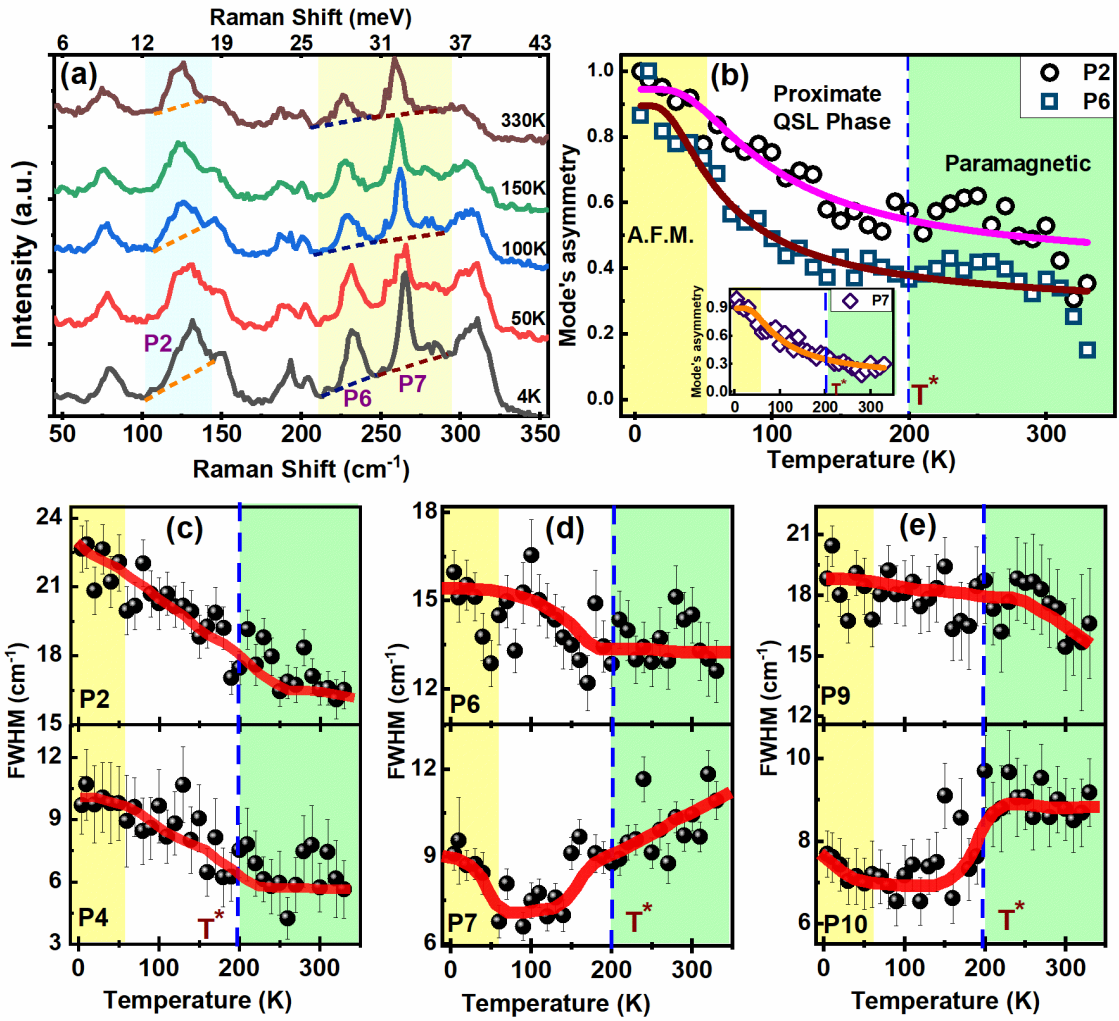

# Supplementary Information

**Fluctuating Fractionalized Spins in Quasi Two-dimensional Magnetic $V_{0.85}PS_3$**

Vivek Kumar[1,*], Deepu Kumar[1], Birender Singh[1], Yuliia Shemerliuk[2,] Mahdi Behnami[2], Bernd Büchner[2,3], Saicharan Aswartham[2], and Pradeep Kumar[1,†]

[1] *School of Basic Sciences, Indian Institute of Technology Mandi, Mandi-175005, India*
[2] *Leibniz-Institute for Solid-state and Materials Research, IFW-Dresden, 01069 Dresden, Germany*
[3] *Institute of Solid-State Physics, TU Dresden, 01069 Dresden, Germany*

[*]vivekvke@gmail.com

[†]pkumar@iitmandi.ac.in

## S1. Synthesis and crystal growth

Chemical vapor transport (CVT) technique was used to grow single crystals of $V_{1-x}PS_3$ with $TeCl_4$ as the transport agent. In a glove box under argon atmosphere, the starting materials of vanadium (powder Alfa Aesar, 99.5%), phosphorus (red, lumps, Alfa Aesar, 99.999%) and sulfur (pieces, Alfa Aesar, 99.999%) were weighed out in a molar ratio of V:P:S = 1:1:3 and thoroughly mixed with 0.1 g of $TeCl_4$ (powder, Alfa Aesar, 99.9%). 1 g of reaction mixture was loaded in a quartz ampule which was then evacuated to a residual pressure of $10^{-8}$ bar and sealed at a length of approximately 12 cm. The closed ampule was placed in a two-zone horizontal tube furnace for CVT growth. To provide a pre-reaction of P and S with the vanadium, the furnace was heated homogeneously to 300°C with 100°C/h and dwelled at this temperature for 24 h. After that, temperature was increased to 600°C at 100°C/h. The charge region was kept at this temperature for 551 h while the sink region was initially heated up to 650°C at 100°C/h, dwelled for 24 h. Then, the sink region was cooled to 350°C for 24 h. The ampule was dwelled with a transport gradient of 600°C (charge) to 350°C (sink) for 504 h. In the last step, the charge region was cooled to the sink temperature in a duration of 1 h and later both sides were cooled to room temperature. A similar approach was used by us for the crystal growth of the closely related sister compounds such as $(Mn_{1-x}Ni_x)_2P_2S_6$, $(Fe_{1-x}Ni_x)_2P_2S_6$ and $AgCrP_2S_6$ [1-3].

Shiny plate-like single crystals of $V_xPS_3$ of up to 2 mm×3 mm×200 μm were obtained as shown in figure S1. All crystals exhibit the typical features of a layered van der Waals structure, and they are easily exfoliated by scotch tape. Figure 1a. shows an example of as-grown single crystal with a well-defined flat facade with 120° angles, which clearly indicates the hexagonal crystal habitus and that the crystals grew along the symmetry axes. As grown crystals were characterized by x-ray diffraction for structural analysis and with Scanning Electron Microscope (SEM) for compositional analysis. Backscattered electron (BSE) images of as grown crystal have shown no change of chemical contrast on the surface of the crystal, as shown in Figure 1b. Some spots of different contrast can be clearly attributed to particles on the surface of the crystal rather than intrinsic impurities by comparing BSE and SE images (see Fig. S1(b-c)). This indicates a homogeneous elemental composition on the respective area of the crystal.

The mean elemental composition of the as-grown single crystals was quantified from energy-dispersive X-ray EDX spectroscopy which was performed on several different areas and points on the surface of multiple crystals of the same crystal growth experiment. The obtained mean elemental composition of $V_{17.(4)}P_{21.0(7)}S_{61.(4)}$, which confirms the off-stoichiometry of vanadium. Coak et.al., [4] explained the vanadium deficiency in $V_{1-x}PS_3$ by valence mixing on the vanadium site between $V^{2+}$ ($d^3$, $S=3/2$) and $V^{3+}$ ($d^2, S=1$) states . The low standard deviations of the mean ratios are exemplary for a homogeneous elemental distribution. We note that in the literature, the general nomenclature used for the members of this family are 113 (i.e., $TMPS_3$) or 226 (i.e., $TM_2P_2S_6$) [2,5-7]. $V_{0.85}PS_3$ crystallizes in monoclinic symmetry, with space group *C*2/m and unit cell parameters are $a = 5.86$ Å, $b = 10.15$ Å, $c = 6.66$ Å, $\alpha = 90.00°$, $\beta = 106.906°$, $\gamma = 90.00°$ and $V = 378.07$ Å$^3$.

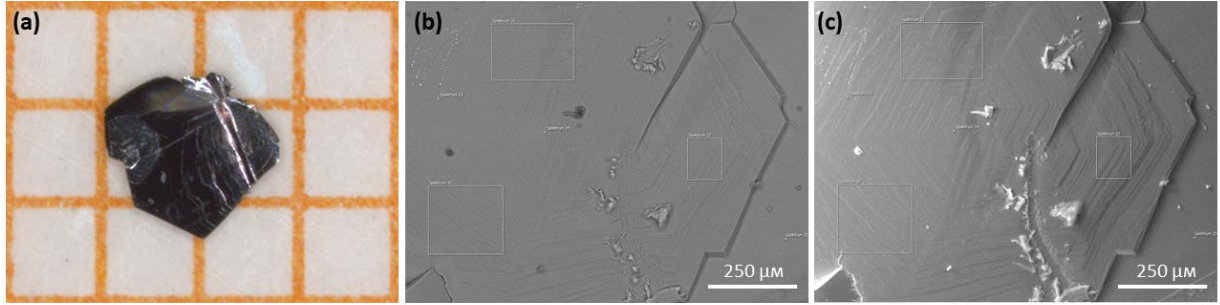

**Figure S1**: (a) Image of as-grown plate-like single crystal grown by the chemical vapor transport. SEM images of a crystal obtained with (b) chemical contrast (BSE detector) and (c) topographical contrast (SE detector)

## S2. Raman Scattering

Inelastically scattered light was collected via micro-Raman spectrometer (LabRAM HR Evolution) in backscattering configuration. Sample was irradiated by a linearly polarized TE-cooled, 532 nm (2.33 eV) laser and focused via 50x LWD objective with N.A. of 0.8. Laser power was kept low (< 0.5 mW) to prevent local heating effect. The scattered light was detected by a Peltier cooled CCD after getting dispersed by a 600 groves/mm grating. Sample is kept on a platform inside a closed chamber with pressure reduced to ~ 1.0 µ Torr. The sample temperature is regulated over a range of 4K - 330 K with ± 0.1 K accuracy using a closed-cycle He-flow cryostat (Montana). Further to unveil the symmetry of phonon modes, we performed polarization-dependent measurement where we varied the polarization direction of the incident light using retarders while the analyzer is kept fixed. Feynman diagram, schematic representation for stokes Raman scattering and schematic experimental setup is shown in figure S2.

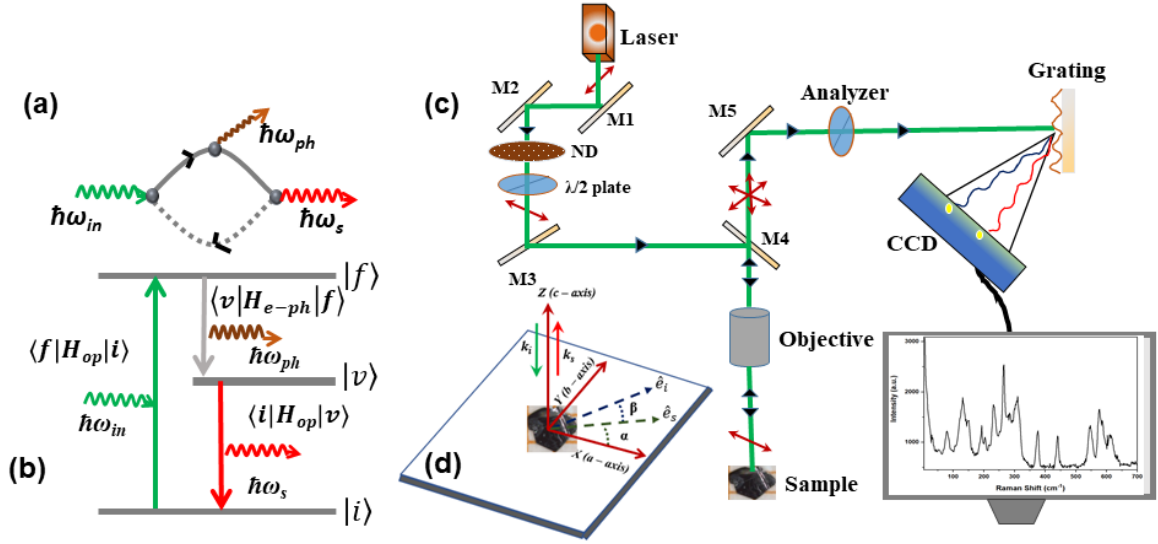

**Figure S2**: **(a)**, **(b)** shows Feynman diagram and schematic representation for the Stokes Raman scattering process, respectively. $\hbar\omega_{in}$, $\hbar\omega_s$ and $\hbar\omega_{ph}$ corresponds to the energy of the incident, scattered photon and created phonon respectively. $|i\rangle$ represents ground state, $|v\rangle$ and $|f\rangle$ are intermediate states. $\langle f|H_{op}|i\rangle$, $\langle v|H_{e-ph}|f\rangle$ and $\langle i|H_{op}|v\rangle$ are the matrix elements for the process of optical absorption, electron-phonon interaction and optical emission. **(c)** Schematic representation of the experimental setup of instrument. **(d)** Plane projection of polarization direction of incident and scattered light.

## S3. Resistivity and magnetization measurements

Fig. S3(a) shows the resistivity vs T plot. Inset shows the log of resistivity as a function of temperature, Arrhenius fitting gives the bandgap of ~ 180 meV. Magnetic susceptibility measurement was performed in Z.F.C. mode at $H_{ex}$ = 1000 Oe shown in Fig. S3(b) shows $T_N$ ~ 60 K respectively, is very close to ($T_N$ ~ 62K) what was reported by Coak et al., [4] Magnetic susceptibility increases slowly as temperature is decreased from 300K and shows a sharp decrease around 60K. In the paramagnetic phase it shows no deviation, but once spin-solid phase is reached the magnetic susceptibility starts to show anisotropic behavior, which suggests that it is comparatively easier to magnetize the sample in *ab* plane than along *c*-axis. Interestingly below ~ 50K, susceptibility again start increasing, which is very unlikely in case of conventional antiferromagnets where net magnetization goes to zero sharply below Néel temperature. Single ion anisotropy (SIA ~ $DS_{iz}^2$) may be a source of anisotropy in Heisenberg case when $J_\perp = J_\parallel$ as also reported for Fe, Mn and Ni case. However, in case of V above $T_N$ (~ 60K) we do see isotropic behavior suggesting negligible SIA effect. For Fe, Mn and Ni, $\chi_{mol}$ *vs T* showed behavior of a typical Anti-ferromagnetic (AFM) system i.e., $\chi_{mol}^\parallel$ parallel to magnetization axis shows sharp drop below $T_N$, whereas $\chi_{mol}^\perp$ perpendicular to the magnetic axis remains nearly constant with slight increase owing to the presence of spin waves[8-10]. Surprisingly in case of $V_{1-x}PS_3$ both $\chi_{mol}^\parallel$ and $\chi_{mol}^\perp$ shows similar (increases with decrease in temperature) behavior below $T_N$ unlike of a typical AFM system. Such a complex magnetic behavior could be a reflection of quantum spin disordered state. In case of SIA if D > 0 then $\chi_{mol}^\perp > \chi_{mol}^\parallel$ and if D < 0 $\chi_{mol}^\perp < \chi_{mol}^\parallel$ , but if D ≈ 0 then $\chi_{mol}^\perp = \chi_{mol}^\parallel$. This was expected for the case of V, because for both $V^{3+}$ ($d^2$ ; S=1) and $V^{2+}$ ($d^3$ ; S=3/2) state $^3A_{2g}$ is the lowest state and is orbitally non-degenerate and it is not expected to be affected by triagonal distortion. Increase in $\chi_{mol}^\parallel$ and $\chi_{mol}^\perp$ below $T_N$ reflect the dominance of only short-range ordering. This can be a signal of a proximate QSL state or may be a remnant of QSL state existing as fluctuation. We note that a similar behavior of $\chi_{mol}$ below $T_N$ was also reported for $Cu_2Te_2O_5Br_2$ [11,12] which could be understood by quantum critical transition between AFM state to QSL state.

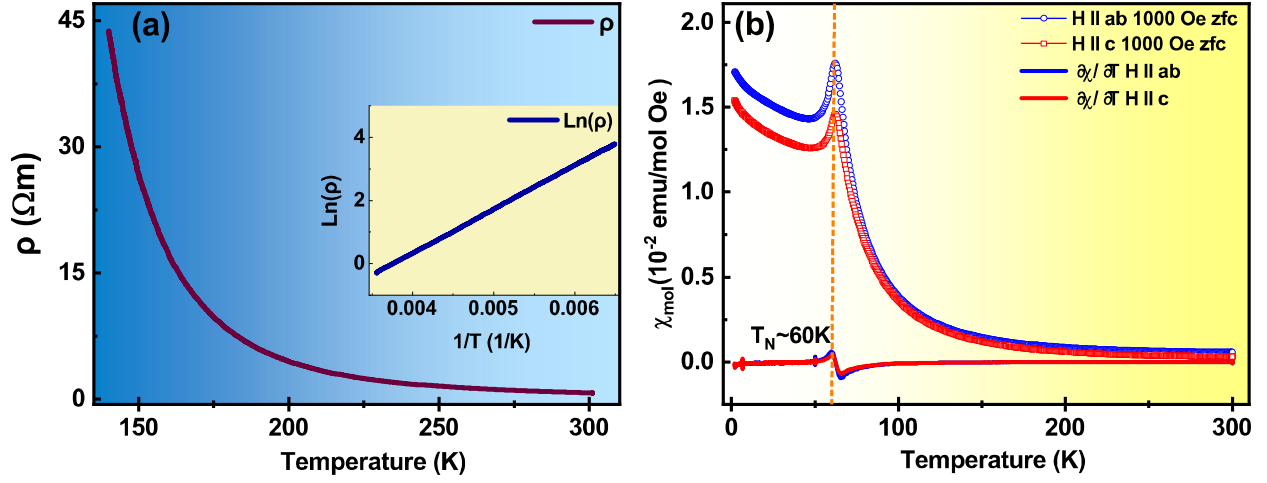

**Figure S3: (a)** Resistivity as a function of temperature and inset shows ln(ρ) vs 1/T, **(b)** Molar magnetic susceptibility $\chi_{mol}$ in Z.F.C. mode with H ∥ ab plane and H ∥ c axis along with respective derivatives.

## S4. Temperature dependent phonons analysis

In the stoichiometric structure of VPS$_3$, the factor group analysis predicts a total of 30 non-degenerate modes, $\Gamma = 8A_g + 7B_g + 6A_u + 9B_u$, within the irreducible representation at the gamma point, out of which 15, $\Gamma_{Raman} = 8A_g + 7B_g$, are Raman active with symmetric $A_g$ and antisymmetric $B_g$ lattice vibrations; and 15, $\Gamma_{infrared} = 6A_u + 9B_u$, are of infrared in nature, details are summarized in Table-SI. Figure S4 shows evolution of the Raman spectra with temperature. We found 15 modes which is consistent with group theory prediction as well. We have fitted the Raman spectra at different temperatures with Lorentzian function and extracted corresponding phonon self-energy parameters. With no new emerging mode, suggesting absence of any structural transitions within the temperature range of 4K – 330 K. In order to understand the underlying physics of temperature dependence of phonon features in case of periodic solids, one may Taylor expand the potential energy about the equilibrium position as:

$$U(q) = U(q_o) + q \left.\frac{\partial U}{\partial q}\right|_{q=q_o} + q^2 \left.\frac{\partial^2 U}{\partial q^2}\right|_{q=q_o} + q^3 \left.\frac{\partial^3 U}{\partial q^3}\right|_{q=q_o} + ... \qquad \text{-- (1)}$$

or equivalently: $\qquad U = const. + U_{harmonic} + U_{anharmonic} \qquad$ -- (2)

$$U_{anharmonic} = gq^3 + mq^4 + ... \equiv \beta(a^+ a^+ a + a^+ aa) + \gamma(a^+ a^+ a^+ a + a^+ a^+ aa + ..) + ... \qquad \text{-- (3)}$$

where '$a^+$' and '$a$' are creation and annihilation operators, q is the normalized coordinate. Within harmonic approximation, gamma point ($\vec{k}=0$) phonon features are independent of temperature. But in reality, anharmonic effects come into the picture. Cubic or higher terms are the source of anharmonicity and because of that vibrational energy levels are no longer equidistant which makes interatomic force constant a function of normal coordinate and consequently is responsible for the shift in frequency of the phonons with temperature and the phonon-phonon interaction is responsible for the finite linewidth of Raman peaks as it is inversely related to the lifetime ($\tau$) of the phonon, $\Gamma(FWHM) \propto \tau^{-1}$. Keeping energy and momentum conservation in check, cubic term ($gq^3$) reflects the three-phonon process where an optical phonon mode decays into two equal energy acoustic phonon modes ($\omega_1 = \omega_2 = \omega/2; k_1 + k_2 = 0$), while quartic term ($mq^4$) indicates four-phonon process which shows decomposition of optical phonon into three identical acoustic phonons ($\omega_1 = \omega_2 = \omega_3 = \omega/3; k_1 + k_2 + k_3 = 0$).

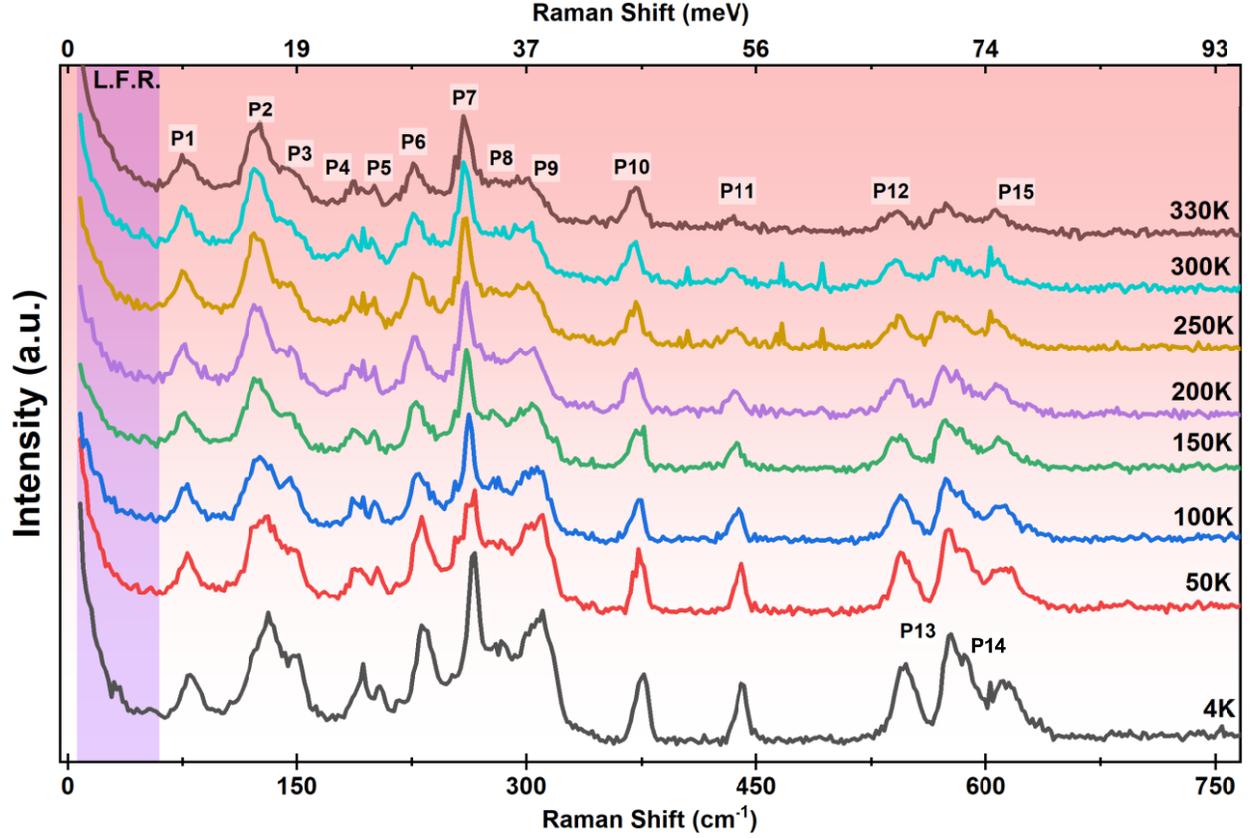

**Figure S4:** Raw Spectra of bulk $V_{0.85}PS_3$ along with peak labels at different temperatures within the spectral range of 5–760 cm$^{-1}$. Low Frequency Region (LFR) is shaded in purple.

Effect of thermal part of anharmonicity can be visualized in temperature dependent variation of phonon frequencies and linewidth in three-phonon process using following functional forms [13]:

$$\omega(T) = \omega_o + A\left(1 + \frac{2}{e^x - 1}\right) \quad \text{--- (4)}$$

$$\Gamma(T) = \Gamma_o + C\left(1 + \frac{2}{e^x - 1}\right) \quad \text{--- (5)}$$

respectively, here $\omega_o$ and $\Gamma_o$ are frequency and line-width at absolute zero; $x = \frac{\hbar\omega_o}{2k_BT}$, $y = \frac{\hbar\omega_o}{3k_BT}$, $A$ and $C$ are the self-energy constants [14]. Three phonon contribution is fitted and shown by thick red curve in figure S5 in a temperature range of 60K to 200K for temperature dependent frequency and linewidth of P1 – P7, P9-P15 modes; and the estimated deviation from cubic anharmonic model below 60K is indicated by an extrapolated dashed red curve. We obtained a negative value of 'A' for all the modes which implies phonon display blueshift with decreasing temperature, which is considered as normal behavior when fitted using cubic anharmonic model, derived parameters are summed up in Table-SII. We clearly spot the deviation from three phonon process as phonon modes blueshifts below transition temperature, which can be attributed to interaction of magnetic and lattice degree of freedom. Interestingly, above 200K we observe temperature independent behavior for most of the phonon modes.

For variation in FWHM, the self-energy constant 'C' is expected to be positive as phonon population decreases with decrease in temperature which increases phonon lifetime and we observe it for modes the P3, P11 and P12, whereas P2, P4 and P6 shows opposite behavior. Interestingly P13 and P14 below 200K remains almost constant till 4K which is anomalous behavior. P2 and P4 show significant increase in FWHM below 200K. Linewidth of the mode P10 shows quite interesting behavior, below 200K it shows sharp drop and remains nearly constant till ~60K and below 60K it starts increasing. Similarly, linewidth of mode P7 shows a drop around 200K and then remains constant till ~ 60K and at lower temperature it starts increasing. Different extent of variation of FWHM for different phonon modes suggest that underlying magnetic degree of freedom are interacting in different fashion with different energy phonon modes.

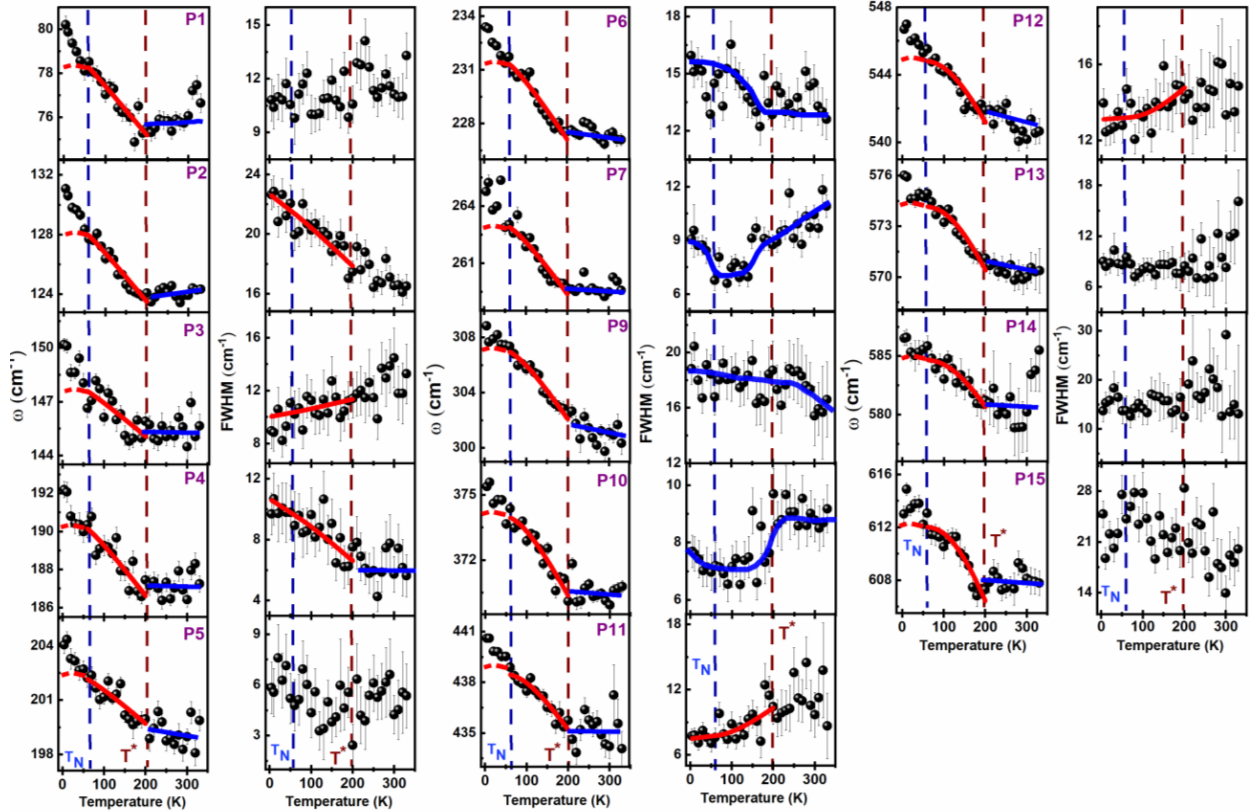

**Figure S5:** Temperature dependent evolution of features (frequency and FWHM) of the phonon modes. Some of phonon were fitted with anharmonic model as mentioned in the text indicated by solid red line and solid blue lines are guide to the eye. Dotted vertical lines represent $T_N$ (~ 60K) and cross-over temperature $T^*$ (~ 200K).

## S5. Spin-phonon coupling

To further understand coupling of lattice degree of freedom with the underlying magnetic degree of freedom, we invoked spin-phonon coupling and tried to understand the phonons behavior at low temperature. As anharmonic effect is more prominent at higher temperatures and at low temperature, below $T_N$ ~ 60K in spin-solid phase due to magnetic interaction another decay channel becomes active, which could be responsible for phonon renormalization. Interestingly, the change in the phonon modes frequency is quite large for some of the modes below $T_N$, for e.g., P1 mode undergoes 3% change in frequency, P2 and P3 ~ 2%. To understand the underlying physics, potential energy for the crystal [15] within harmonic approximation in presence of spin-spin correlation may be given as :

$$U(q) = U_o + \sum_{i, i \neq j} J_{ij} \vec{S}_i \cdot \vec{S}_j \qquad \text{--- (6)}$$

where $U_o = \frac{1}{2}\xi q^2$, here $\xi$ and $q$ are the force constant and displacement from equilibrium position respectively, $J_{ij}$ is the exchange coupling constant and $\vec{S}_i \cdot \vec{S}_j$ is spin-spin correlation between $i^{th}$ and $j^{th}$ magnetic ion. As atoms are in motion, second derivative of potential is given as:

$$\frac{\partial^2 U}{\partial q^2} = \xi + \sum_{i, i \neq j} \frac{\partial^2 J_{ij}}{\partial q^2} \langle \vec{S}_i \cdot \vec{S}_j \rangle \qquad \text{--- (7)}$$

where the first term is the force constant pertaining to the harmonic part and second term arises due to variation in exchange coupling constant subjected to lattice vibration. So, this additional renormalization of phonon mode frequency arises due to coupling of lattice and magnetic degree of freedom in the long range ordered phase. In such a case the renormalized phonon frequency takes the following form:

$$\Delta \omega \approx \omega_{sp} - \omega_o^{ph}(T) = \lambda \langle \vec{S}_i \cdot \vec{S}_j \rangle \qquad \text{--- (8)}$$

where $\omega_{sp}$, $\omega_o^{ph}(T)$ correspond to phonon mode frequency and bare phonon frequency (without spin-phonon coupling), respectively, and $\lambda = \frac{\partial^2 J_{ij}}{\partial q^2}$ is the spin-phonon coupling constant [16-18]. So, we fitted the frequency variation with temperature below $T_N$ for different modes using following relation of spin-phonon coupling [19],

$$\Delta\omega \approx \omega_{sp} - \omega_o^{ph}(T) = -\lambda * S^2 * \phi(T) \qquad \text{--- (9)}$$

where $S$ is spin on the magnetic ion and $\phi(T)$ is the order parameter given as

$\phi(T) = 1 - \left(\dfrac{T}{T_N^*}\right)^{\gamma}$ ; $\gamma$ is a critical exponent. We have kept $T_N^*$ as a variable and the obtained value is close to the value obtained from magnetic measurements. Due to presence of mixed valency, the value of 'S' is taken as a weighted average of spin on $V^{2+}$ and $V^{3+}$: $S = \dfrac{0.30*1 + 0.55*1.5}{0.30 + 0.55} = 1.32$; derived parameters are tabulated in Table-SIII.

## S6. Polarization dependent study

To understand symmetry of the phonon modes. We employed angle resolved polarized Raman scattering experiment, which can be performed in different equivalent configurations for e.g., rotating incident light polarization direction, scattered light and the sample itself. We did our experiment by varying the incident light polarization direction $20^o$ from $0^o$ to $360^o$ using a half wave retarder or $\lambda/2$ plate which rotates incident light polarization direction to twice the angle which incident light polarization makes with the optical axis of $\lambda/2$ plate, whereas analyzer has been kept fixed. Intensity of the inelastically scattered light can be written as:

$$I_{Raman} \propto \left|\hat{e}_s^t . R . \hat{e}_i\right|^2 \qquad \text{--- (10)}$$

where '$t$' symbolizes transpose, '$\hat{e}_i$' and '$\hat{e}_s$' are the unit vectors of incident and scattered light polarization direction. '$R$' represents the Raman tensor of the respective phonon mode [20-22]. A schematic diagram of polarization vectors of incident and scattered light projected on a plane is shown in fig. S2 (d). In the matrix form, incident and scattered light polarization direction unit vector can be written as: $\hat{e}_i = [\cos(\alpha + \beta) \ \sin(\alpha + \beta) \ 0]$ ; $\hat{e}_s = [\cos(\alpha) \ \sin(\alpha) \ 0]$, where '$\beta$' is the relative angle between '$\hat{e}_i$' and '$\hat{e}_s$' and '$\alpha$' is an angle of scattered light from x-axis, when polarization unit vectors are projected in x (*a*-axis) - y (*b*-axis) plane. The angle dependency of intensities of $A_g$ and $B_g$ modes using Raman tensor mentioned in Table-SI can be written as:

$$I_{A_g} = |a\cos(\alpha)\cos(\alpha+\beta) + b\sin(\alpha)\sin(\alpha+\beta)|^2 \qquad \text{--- (11)}$$

$$I_{B_g} = |e\cos(\alpha)\sin(\alpha+\beta) + e\sin(\alpha)\cos(\alpha+\beta)|^2 \qquad \text{--- (12)}$$

which are expected to have a phase difference of π/2 in intensity variation. Here $\alpha$ is an arbitrary angle from the *a*-axis and is kept constant. Therefore, without any loss of generality it is taken as zero, giving rise to the expression for the Raman intensity as $I_{A_g} = |a\cos(\beta)|^2$ and $I_{B_g} = |e\sin(\beta)|^2$. Polarization dependence of the phonon modes and evolution of raw spectrum below and above $T_N$ is shown in fig. S6 and fig. S7 respectively. The high energy modes P6-P10, P12, P14-P15 shows typical of an $A_g$ symmetry mode intensity consistent with group theoretical predictions i.e., two maxima at zero and $\pi$, the red lines are the fitted curve using $I_{A_g} = |a\cos(\beta)|^2$ and the fitting is quite good. However, the modes P1-P5, P11 and P13 have elongated dumble shape with maxima around $\pi/4$ and $3\pi/4$ suggesting $B_g$ symmetry. The tentative symmetry assignment for the observed phonon modes is given in Table-SII.

We did polarization dependent measurements below and above long-range magnetic ordering i.e., at 4K and 120K. Quite interestingly and fascinating observation is that for some of the modes the major axis, the axis along which we see maxima in the intensity as a function of angle, is rotated in the low temperature magnetic phase (4K) as compared to the non-magnetic phase (120K). For e.g., P13 modes major axis at 120K (above $T_N$) is at ~ 0 and is shifted to ~ $\pi/4$ at 4K (below $T_N$). For modes P4 and P5 are quasi-isotropic at 120K and the major axis at 4K is at ~ $\pi/4$. For mode P1 major axis above at 120K is at ~ $\pi/2$ and is shifted to ~$\pi/4$ at 4K. Such a rotation of major axis suggests the possibility of the tunability of the scatted light in this quasi 2D magnet via symmetry control. As one enters the spin-solid phase both time reversal and spin rotational symmetries are broken and these broken symmetries does have impact on the underlying spin degrees of freedom possibly resulting into magneto-optical Raman effect, reflected via rotating the scattered light by ~ 30-45 degrees. Our results highlight the possibility to control the quantum pathways of the inelastically scattered light in these 2D magnetic systems and this control may provide a way in future to manipulate these pathways for use in quantum technology.

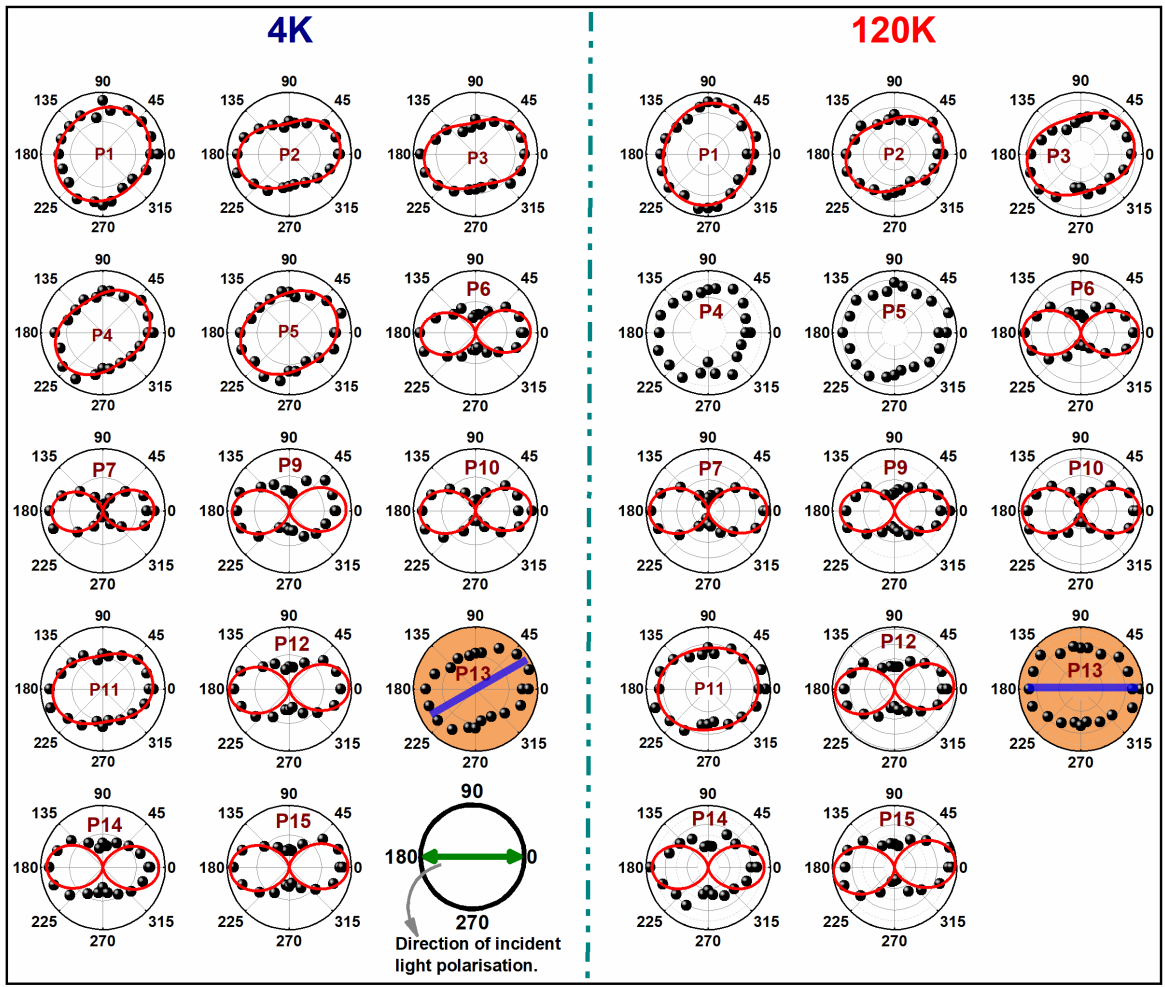

**Figure S6:** Polarization dependent intensity of phonon modes at 4K and 120K.

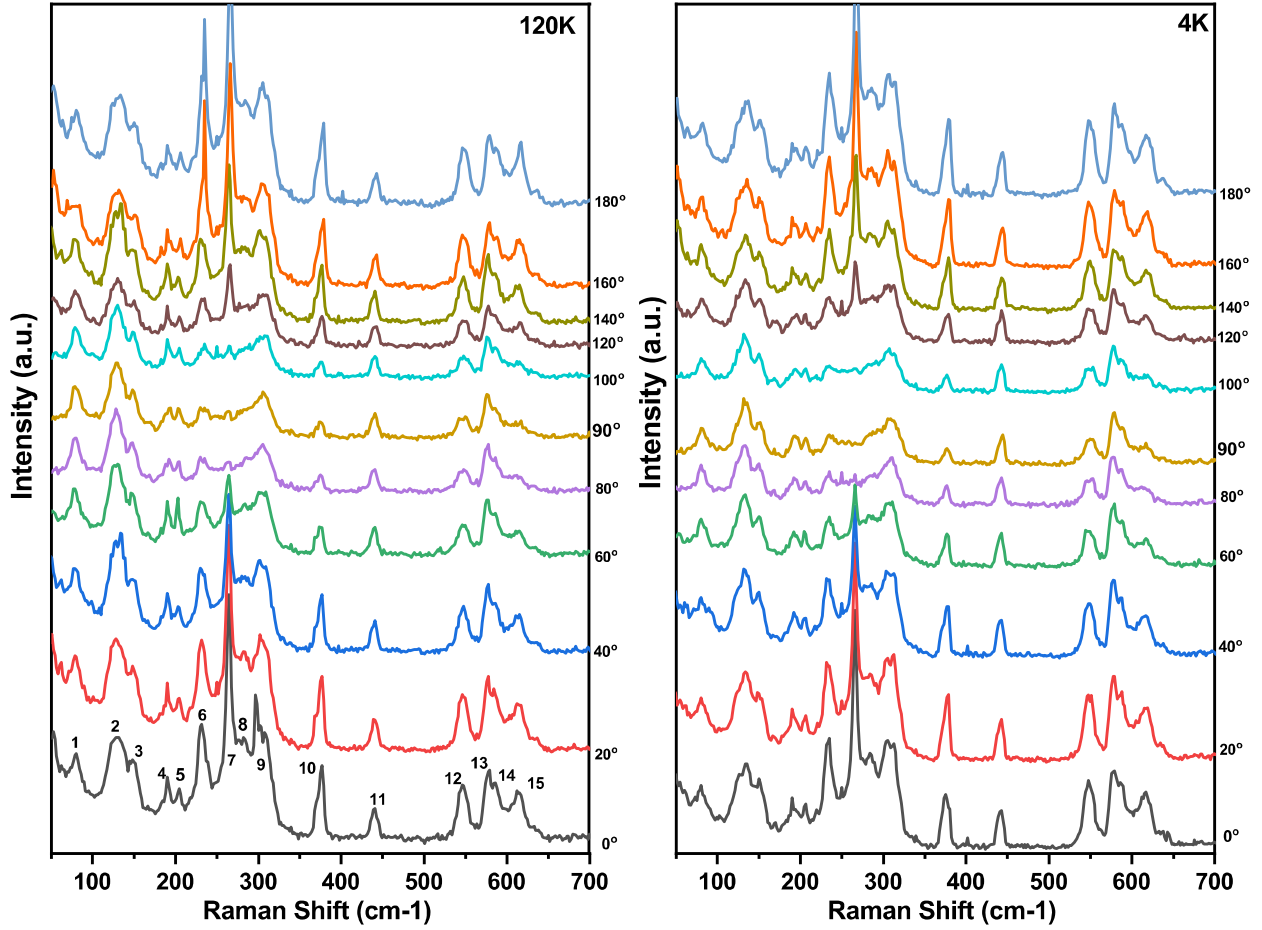

**Figure S7:** Polarization dependent evolution of the spectrum in the low temperature magnetic phase (4K) and at high temperature non-magnetic phase (120K).

## S7. Calculation of phonon asymmetry using slope method

Here, we have employed slope method to find the nature of the phonon modes asymmetry as the Fano function fitting gives rise to large error. The slope ($\Delta Y / \Delta X = \Delta Inten. / \Delta \omega$) for a mode is evaluated by keeping the x-axis range (frequency here) same and the corresponding rise in the intensity as a function of temperature. Fano function is defined as $F(\omega) = I_0 (q + \varepsilon)^2 / (1 + \varepsilon^2)$; where $\varepsilon = (\omega - \omega_0) / \Gamma$ and $1/q$ defines as the asymmetry. The asymmetry parameter ($1/q$) characterizes the coupling strength of a phonon to the underlying continuum: a stronger coupling ($1/q \to \infty$) causes the peak to be more asymmetric and in the weak-coupling limit ($1/q \to 0$) the Fano line shape is reduced to a Lorentzian line shape. We generated this function by keeping $\omega_0$, $\Gamma$ and $I_0$ fixed for different values of $q$, please see Fig. S8 (a), and it's very clear from this figure that as $1/q$ increases asymmetry is increased. Then, we estimated the slope as defined above for this generated function and found the one-to-one correspondence between slope and asymmetry parameter $1/q$, please see Fig. 8(b), higher the slope more the asymmetry and lesser the slope lesser the asymmetry.

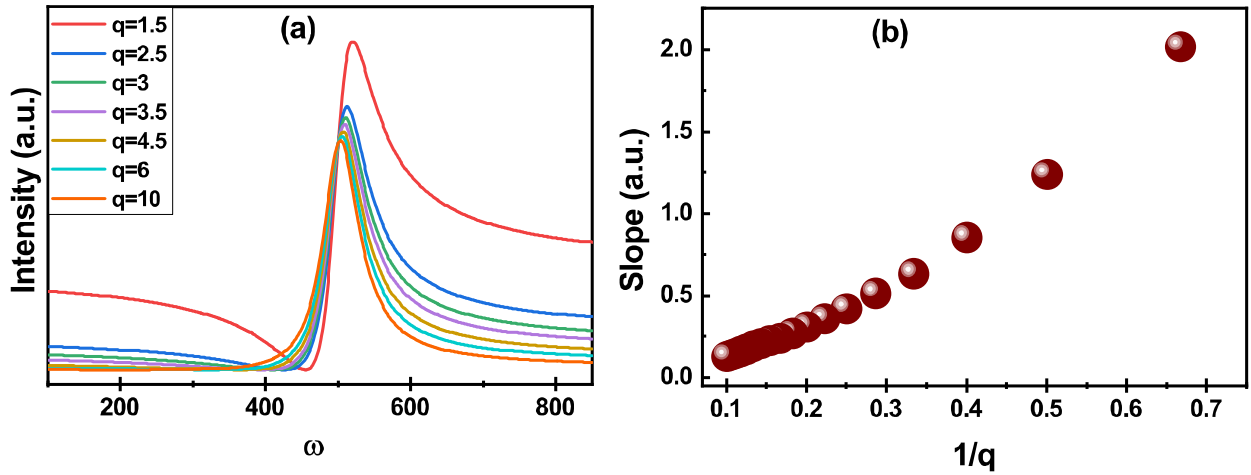

**Figure S8**: (a) Evolution of a mode for different $q$ parameter, evolution shows high asymmetry at low $q$ value to close to Lorentzian functional form at high $q$ value. (b) Plot of slope v/s asymmetry parameter (*1/q*).

**Table-SI:** Wyckoff positions of different atoms in conventional unit cell and irreducible representations of the phonon modes of monoclinic (C2/m [$C_{2h}$] $V_{1-x}PS_3$) at the gamma point. $R_{A_g}$ and $R_{B_g}$ are the Raman tensors for $A_g$ and $B_g$ phonon modes.

| Atoms | Wyckoff site | Γ-point mode decomposition | Raman Tensor |
|---|---|---|---|
| V | 4g | $A_g + A_u + 2B_g + 2B_g$ | $R_{A_g} = \begin{pmatrix} a & 0 & d \\ 0 & b & 0 \\ d & 0 & c \end{pmatrix}$ |
| P | 4i | $2A_g + A_u + B_g + 2B_u$ | |
| S | 4i | $2A_g + A_u + B_g + 2B_u$ | $R_{B_g} = \begin{pmatrix} 0 & e & 0 \\ e & 0 & f \\ 0 & f & 0 \end{pmatrix}$ |
| S | 8j | $3A_g + 3A_u + 3B_g + 3B_u$ | |
| | $\Gamma_{Raman} = 8A_g + 7B_g$ | $\Gamma_{Infrared} = 6A_u + 9B_u$ | |

**Table-SII:** Anharmonic fitting parameters of the phonon modes and symmetry assignment.

| Modes (Symmetry) | $\omega_o$ | A | $\Gamma_o$ | C |
|---|---|---|---|---|
| P1 (Bg) | 79.7 ± 0.4 | - 0.6 ± 0.1 | | |
| P2 (Bg) | 130.4 ± 0.4 | - 1.6 ± 0.1 | 23.3 ± 0.7 | - 1.2 ± 0.2 |
| P3 (Bg) | 149.1 ± 0.7 | - 1.1 ± 0.2 | 9.8 ± 1.1 | 0.4 ± 1.4 |
| P4 (Bg) | 192.5 ± 0.7 | - 1.9 ± 0.3 | 11.7 ± 1.1 | - 1.7 ± 0.5 |
| P5 (Bg) | 203.9 ± 0.5 | - 1.5 ± 0.2 | | |
| P6 (Ag) | 234.7 ± 0.4 | - 3.1 ± 0.2 | | |
| P7 (Ag) | 266.2 ± 0.4 | - 3.0 ± 0.2 | | |
| P9 (Ag) | 312.5 ± 0.5 | -5.3 ± 0.4 | | |
| P10 (Ag) | 379.5 ± 0.6 | - 5.4 ± 0.5 | | |
| P11 (Bg) | 445.1 ± 0.8 | - 6.6 ± 0.7 | 2.5 ± 2.3 | 5.2 ± 2.0 |
| P12 (Ag) | 556.0 ± 1.5 | - 11.7 ± 1.4 | 8.0 ± 2.8 | 5.1 ± 2.6 |
| P13 (Bg) | 588.0 ± 1.6 | - 13.9 ± 1.5 | | |
| P14 (Ag) | 600.1 ± 2.5 | - 15.4 ± 2.3 | | |
| P15 (Ag) | 637.1 ± 3.1 | - 25.0 ± 2.9 | | |

**Table-SIII**: Spin-Phonon coupling fitting parameters for some of the prominent phonon modes.

| | Spin-Phonon Coupling | | |
|---|---|---|---|
| | $\lambda$ | $\Upsilon$ | $T_N^*$ |
| **P1:** | -2.0 ± 0.9 | 0.3 ± 0.3 | 69.8 ± 08.8 |
| **P2:** | -2.1 ± 0.2 | 0.9 ± 0.2 | 67.0 ± 03.5 |
| **P3:** | -1.8 ± 0.5 | 1.0 ± 0.7 | 65.8 ± 09.1 |
| **P5:** | -1.4 ± 0.6 | 0.5 ± 0.4 | 60.6 ± 07.4 |
| **P6:** | -1.7 ± 0.4 | 0.6 ± 0.3 | 75.8 ± 09.3 |
| **P7:** | -1.3 ± 0.3 | 1.9 ± 1.8 | 68.7 ± 10.6 |
| **P9:** | -1.3 ± 1.2 | 0.4 ± 0.7 | 93.9 ± 49.7 |
| **P10:** | -1.2 ± 0.3 | 0.9 ± 0.7 | 76.8 ± 17.0 |
| **P11:** | -1.3 ± 0.1 | 1.1 ± 0.4 | 74.1 ± 06.3 |
| **P12:** | -1.2 ± 0.2 | 0.9 ± 0.5 | 75.3 ± 10.6 |